\documentclass[twocolumn,showpacs,preprintnumbers,amsmath,amssymb]{revtex4}
\usepackage{graphicx}
\usepackage{dcolumn}
\usepackage{bm}

\newcommand{\bef}{\begin{figure}}
\newcommand{\eef}{\end{figure}}

\newcommand{\be}{\begin{equation}}
\newcommand{\ee}{\end{equation}}
\newcommand{\bea}{\begin{eqnarray}}
\newcommand{\eea}{\end{eqnarray}}
\newcommand {\mean}[1]{\left\langle #1 \right\rangle}

\newcommand {\mom}   {\mbox{\rm GeV$\kern-0.15em /\kern-0.12em c$}}
\newcommand {\mmom}  {\mbox{\rm MeV$\kern-0.15em /\kern-0.12em c$}}
\newcommand {\mass}  {\mbox{\rm GeV$\kern-0.15em /\kern-0.12em c^2$}}
\newcommand {\mmass} {\mbox{\rm MeV$\kern-0.15em /\kern-0.12em c^2$}}

\newcommand {\tria}{\varepsilon_{3}}
\newcommand {\ecc}{\varepsilon_{2}}

\DeclareMathOperator{\atantwo}{atan2}

\begin{document}

\title{Elliptic and Triangular flow in asymmetric heavy-ion collisions}

\author{Md. Rihan Haque, Md. Nasim, and Bedangadas Mohanty}
\affiliation{Variable Energy Cyclotron Centre, Kolkata 700064, India}

\date{\today}
\begin{abstract}
We present a study of the elliptic ($v_{2}$) and triangular ($v_{3}$) flow
and their corresponding eccentricity fluctuations for asymmetric 
(Au+Ag, Au+Cu and Au+Si) collisions  at $\sqrt{s_{\mathrm {NN}}}$ = 200 GeV.  
These are compared to the corresponding results from symmetric (Au+Au and Cu+Cu) 
collisions at the same energy. The study which is carried out using a multi-phase 
transport (AMPT) model shows that triangularity ($\epsilon_{3}$), fluctuations in triangularity 
and $v_{3}$ do not show much variation for the different colliding ion sizes studied. 
However the eccentricity ($\epsilon_{2}$), fluctuations in eccentricity and 
$v_{2}$ shows a strong dependence on colliding ion size for a given number of
participating nucleons. 
Our study thus indicates that asymmetric heavy-ion collisions could be used to 
constrain models dealing with flow fluctuations in heavy-ion collisions.
\end{abstract}
\pacs{25.75.Ld}
\maketitle

\section{INTRODUCTION}
Knowing the initial geometry and fluctuations in heavy-ion collisions has recently
been shown to have important consequences on interpretating the data from the 
Relativistic Heavy Ion Collider (RHIC) and the Large Hadron Collider (LHC) experiments~\cite{alver,alice}.
For example, the long standing physical interpretation of dihadron
correlations in azimuthal angle~\cite{mach} and pseudorapidity~\cite{ridge} to be due to propagation of
shock waves (due to high momentum quarks or gluons) in the medium and 
jet-medium interactions are being revisited~\cite{alver,alice2}. 
The contribution from the odd harmonics associated with the particle azimuthal angle
distribution (originally thought to be zero due the left-right symmetry in the transverse 
plane of symmetric heavy-ion collisions) to dihadron correlations are found to be important. 
Now it is being widely discussed that non zero 
odd harmonic contributions could arise from fluctuations in the transverse positions
of nucleons undergoing hadronic scattering. Further it is now known that even the 
ideal hydrodynamics calculations under predicts the measured elliptic flow in central heavy-ion
collisions at RHIC~\cite{v2flucth}. Only inclusion of flow fluctuations arising due to the eccentricity fluctuations
can provide some satisfactory explanation of the data. It is also believed that measurement
of higher order flow coefficients and their fluctuations can substantially improve the
constraints on the transport properties of the system formed in high energy heavy-ion collisions.

Experimentally it has not been possible to separate the contribution from elliptic flow ($v_{\rm 2}$),
fluctuations and non-flow from the data~\cite{flowflucnonflow}. This is due to the lack of knowledge of the 
probability distributions for flow fluctuations. The fluctuations in $v_{\rm 2}$ can arise due to
fluctuations in the eccentricity of the overlap region of the two colliding nuclei or deviations 
of the participant plane from the reaction plane. While non-flow effects are those correlations among 
particles that are not related to the reaction plane (e.g due to resonance decay and jets).

Asymmetric heavy-ion collisions could provide density profiles that are different or not
accessible through symmetric heavy-ion collisions. For example a mid-central Cu on Au collision
could lead to the Cu nucleus being occluded in the Au, leading to non zero odd harmonic of 
flow and being highly sensitive to early time dynamics. In addition asymmetric heavy-ion
collisions could provide opportunity to disentangle the path length and energy dependence
of parton energy loss. While in symmetric systems both initial energy density and transverse
size of the medium increases with collision centrality, for asymmetric heavy-ion collisions
it is possible to encounter situations where the transverse dimension of the medium is same
but one can get variation in energy density. Keeping these aspects in mind, RHIC has now 
proposed to carry out asymmetric heavy-ion collisions program~\cite{pac}. In this report we only concentrate
on the study of the variation of $\epsilon_{\rm 2}$, $\epsilon_{\rm 3}$, their fluctuations, $v_{\rm 2}$
and $v_{\rm 3}$ for Au+Au, Au+Ag, Au+Cu, Cu+Cu and Au+Si collisions at $\sqrt{s_{\mathrm {NN}}}$ 
= 200 GeV using a multi-phase transport model (AMPT) with default settings. The goal being 
to simulate the expectation  of above observables in asymmetric heavy-ion collisions relative to
symmetric heavy-ion collisions.

The paper is organized as follows. The next section deals with the definition of 
$\epsilon_{\rm 2}$, $\epsilon_{\rm 3}$, their fluctuations, $v_{\rm 2}$ and $v_{\rm 3}$, 
along with a brief description of AMPT model. Section III presents the results from the
model calculation for the observables mentioned above. Finally we summarize our findings in
section IV.

\section{AMPT model and definitions}

The current work is based on the AMPT model with default settings~\cite{ampt}. It uses the same initial conditions as 
considered in the Heavy Ion Jet Interaction Generator (HIJING) event generator~\cite{hijing}.
HIJING is a perturbative QCD inspired model which produces multiple minijet partons, 
these later get transformed into string configurations and then fragment to hadrons based
on the Lund jet fragmentation model~\cite{lund}. A Glauber model prescription is followed for
obtaining the number of participating nucleons ($N_{\rm{ part}}$). In AMPT, the minijet partons are made 
to undergo scattering before they are allowed to fragment into hadrons. These interactions
could give rise to anisotropy in particle production along azimuthal direction. The event
plane in AMPT is along the x-axis. There exists another version of the model, not used
in this paper, called the string melting version where partonic interactions are considered~\cite{ampt}.

We have followed the notations for the various observables as studied in Ref~\cite{alver}. 
The participant eccentricity is defined as:

\begin{equation}
  \ecc = \frac{\sqrt{\mean{r^2\cos(2\phi_{\text{part}})}^2 + \mean{r^2\sin(2\phi_{\text{part}})}^2}}{\mean{r^2}}
\label{eq:ecc}
\end{equation}
where $r$ and $\phi_{\text{part}}$ are the polar coordinate positions of participating nucleons in the
AMPT model.  $\psi_{2}$ is the angle of the minor axis of the ellipse defined by this region and is given as
\begin{equation}
  \psi_{2}=\frac{\atantwo\left(\mean{r^2\sin(2\phi_{\text{part}})},\mean{r^2\cos(2\phi_{\text{part}})}\right)
+\pi}{2}.
\label{eq:phiecc}
\end{equation}
$v_2$ which is the 2nd Fourier coefficient of the particle distribution 
with respect to $\psi_{2}$ and is given by
\begin{equation}
v_2 = \mean{\cos(2(\phi-\psi_2))}.
\label{eq:v2}
\end{equation}

Similar to the definition of the eccentricity and elliptic flow, the
participant triangularity, $\tria$, and triangular flow, $v_3$ are defined as:
\begin{eqnarray}
  \tria &=& \frac{\sqrt{\mean{r^2\cos(3\phi_{\text{part}})}^2 + \mean{r^2\sin(3\phi_{\text{part}})}^2}}{\mean{r^2}} \label{eq:tria} \\
v_3 &=& \mean{\cos(3(\phi-\psi_3))}
\label{eq:v3}
\end{eqnarray}
 where $\psi_3$ is the angle of the minor axis of participant triangularity and is given by
 \begin{equation}
   \psi_{3}=\frac{\atantwo\left(\mean{r^2\sin(3\phi_{\text{part}})},\mean{r^2\cos(3\phi_{\text{part}})}\right)
     +\pi}{3}.
\label{eq:phitria}
\end{equation}

\section{RESULTS}

\bef
\begin{center}
\includegraphics[scale=0.32]{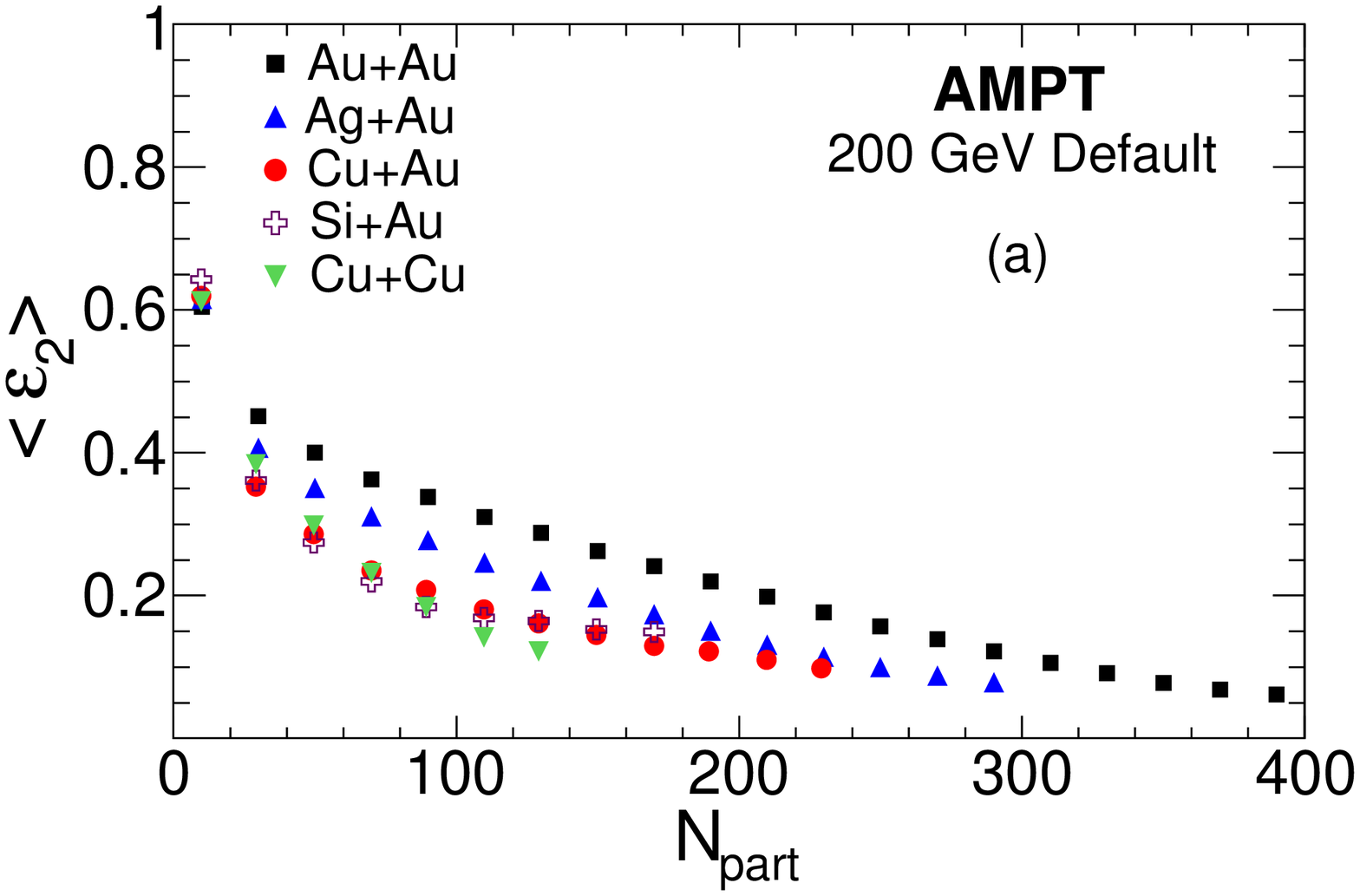}
\includegraphics[scale=0.32]{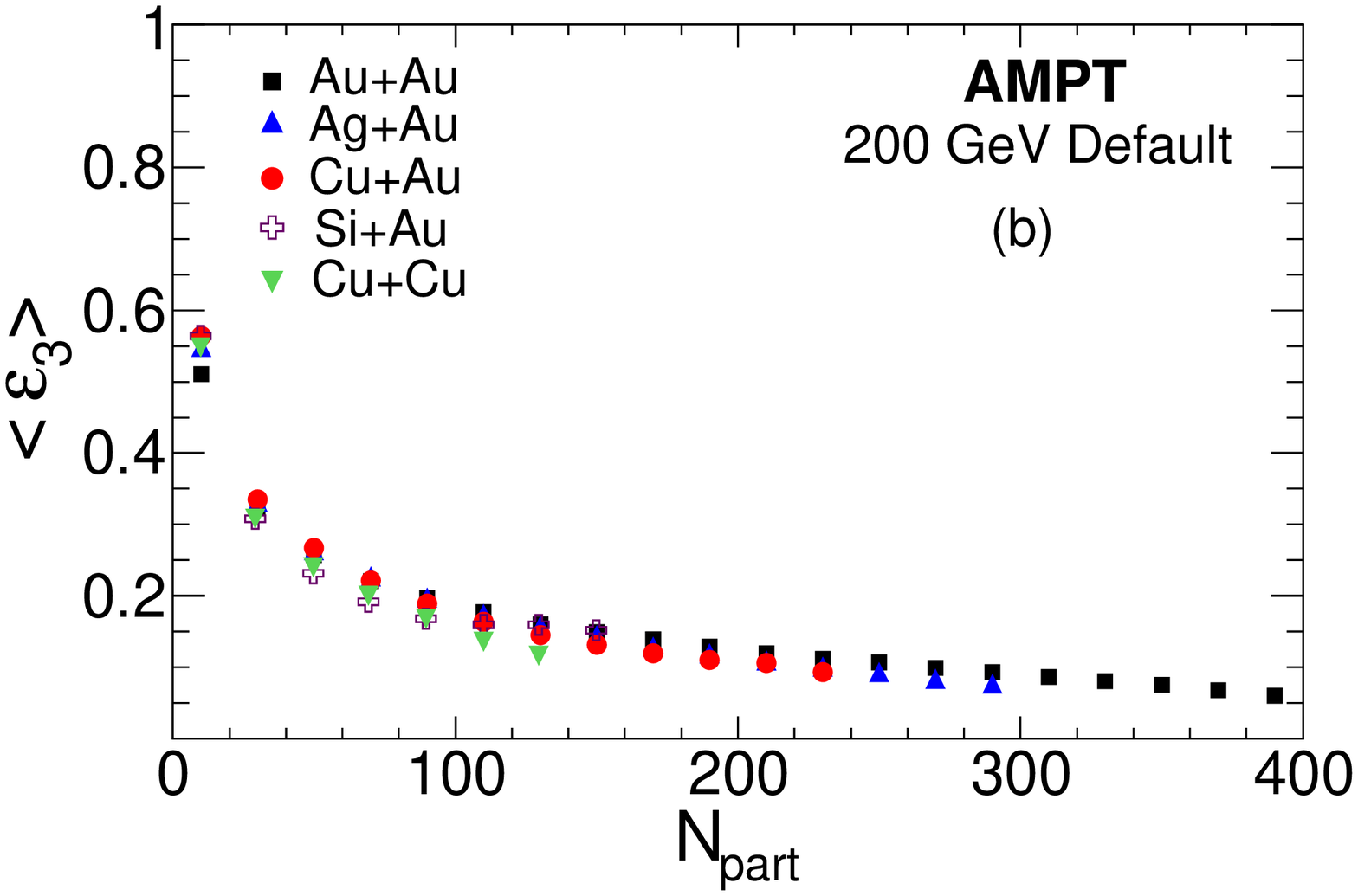}
\caption{(Color online) (a) Average eccentricity ($\langle \ecc \rangle$) 
and (b) average triangularity ($\langle \tria \rangle$) 
as a function of $N_{\rm {part}}$) for the various heavy-ion collisions 
at $\sqrt{s_{\rm {NN}}}$ = 200 GeV from AMPT model.}
\label{fig1}
\end{center}
\eef

Figure~\ref{fig1} shows the $\langle \ecc \rangle$ and $\langle \tria \rangle$ for asymmetric heavy ion 
collisions (Ag+Au, Cu+Au and Si+Au) compared to symmetric heavy-ion collisions (Au+Au and Cu+Cu) 
at $\sqrt{s_{\rm {NN}}}$ = 200 GeV as a function of $N_{\rm {part}}$.
We find for a given $N_{\rm {part}}$, the $\langle \ecc \rangle$ values are higher for larger colliding 
ion sizes. However for the low mass number colliding ions (Cu+Cu collisions and smaller) they are similar.
No such large differences are observed for $\langle \tria \rangle$. This suggests that within the framework 
of the AMPT model, the asymmetric heavy ion collisions can be used to constrain the models 
dealing with second harmonic flow coefficient and its fluctuations. However, such collisions  may not be 
that sensitive to studies dealing directly with triangularity and triangular flow.

\bef
\begin{center}
\includegraphics[scale=0.3]{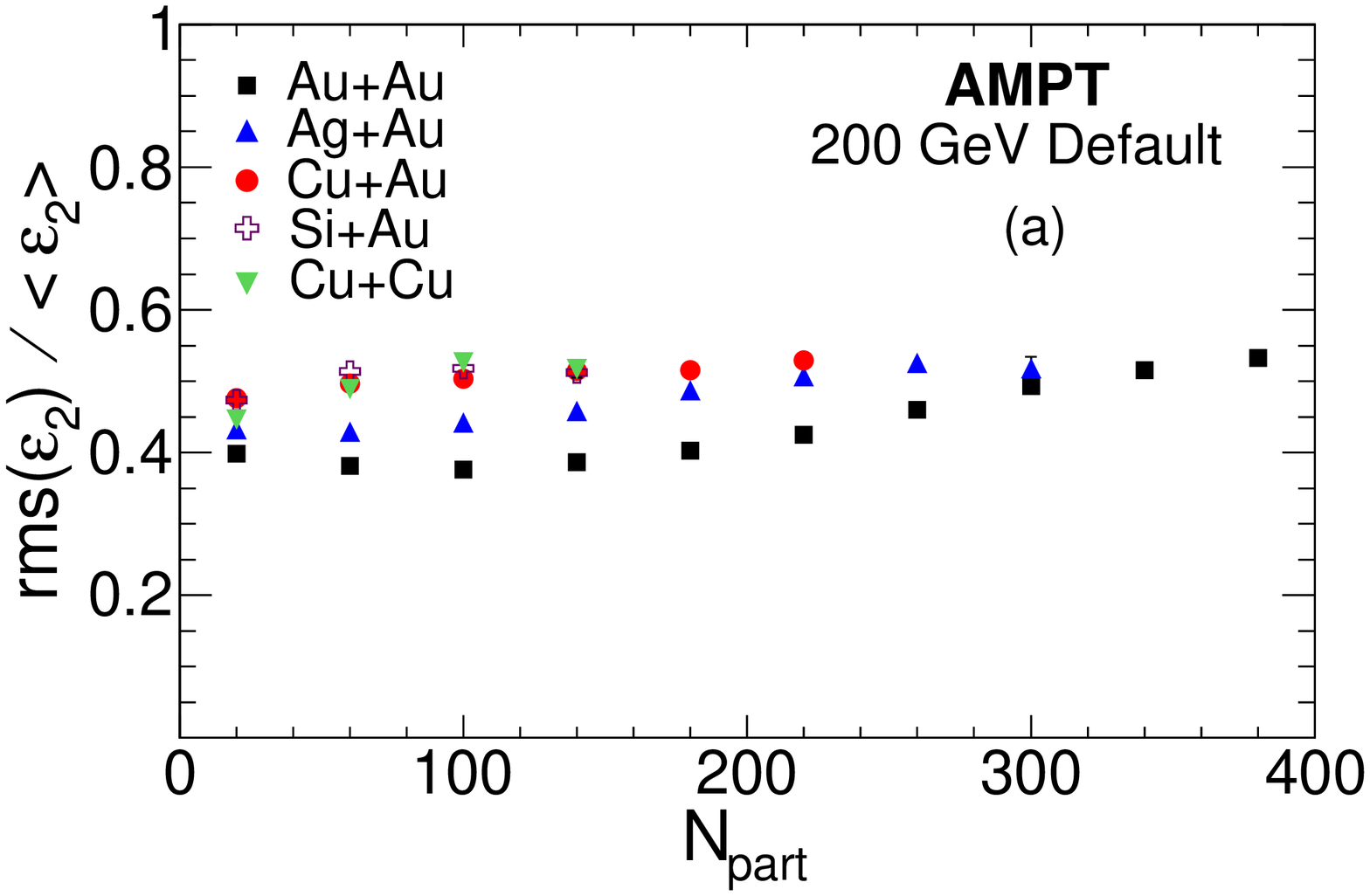}
\includegraphics[scale=0.3]{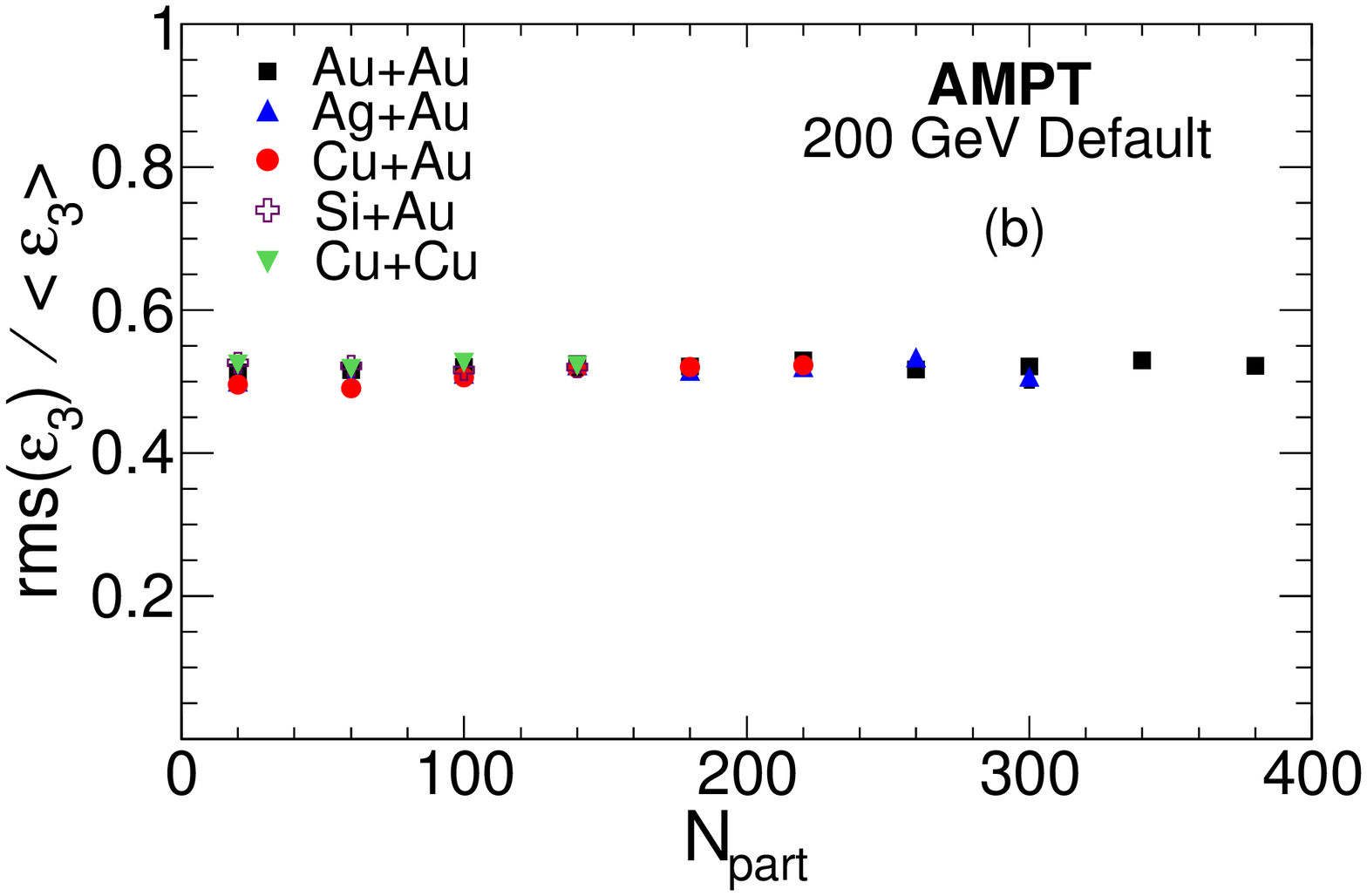}
\caption{(Color online) (a) Ratio of root mean square (rms) value of $\ecc$ to $\langle \ecc \rangle$ and (b) rms of $\tria$ to 
$\langle \tria \rangle$ for various heavy ion collisions at $\sqrt{s_{\rm {NN}}}$ = 200 GeV using AMPT model.}
\label{fig2}
\end{center}
\eef
Figure~\ref{fig2} shows the fluctuations in $\ecc$ and $\tria$ expressed as the ratio of 
the corresponding root mean square (rms) values to their average values for both asymmetric
heavy-ion collisions and symmetric heavy-ion collisions at $\sqrt{s_{\rm {NN}}}$ = 200 GeV
as a function of $N_{part}$. We observe that for mid-central collisions (60 $<$ $N_{\rm {part}}$ $<$ 250)
the fluctuations in $\ecc$ increases as the system size decreases and then saturates for colliding
systems of size comparable or smaller to Cu+Cu. In contrast the variation in fluctuation of $\tria$ 
as a function of system size is very small. 

Similar conclusions are obtained using a slightly 
different observable as proposed in Ref.~\cite{bhalerao}. The results of which from the AMPT model 
are shown in Fig.~\ref{fig3}. It has been shown that the relative magnitude of $v_{\rm n}\{4\}$ and $v_{\rm n}\{2\}$ 
depends on the event-by-event fluctuations of $v_{\rm n}$ if the non flow effects are small. Where the $\{4\}$ and 
$\{2\}$ represents the four particle and two particle cumulant methods to extract flow coefficients
respectively. It has been shown in Ref.~\cite{bhalerao} that 
\begin{equation}
\label{4cumulant}
\left(\frac{v_n\{4\}}{v_n\{2\}}\right)^4=
\left(\frac{\varepsilon_n\{4\}}{\varepsilon_n\{2\}}\right)^4\equiv 
2-\frac{\langle\varepsilon_n^4\rangle}{\langle\varepsilon_n^2\rangle^2}. 
\end{equation} 
\bef
\begin{center}
\includegraphics[scale=0.32]{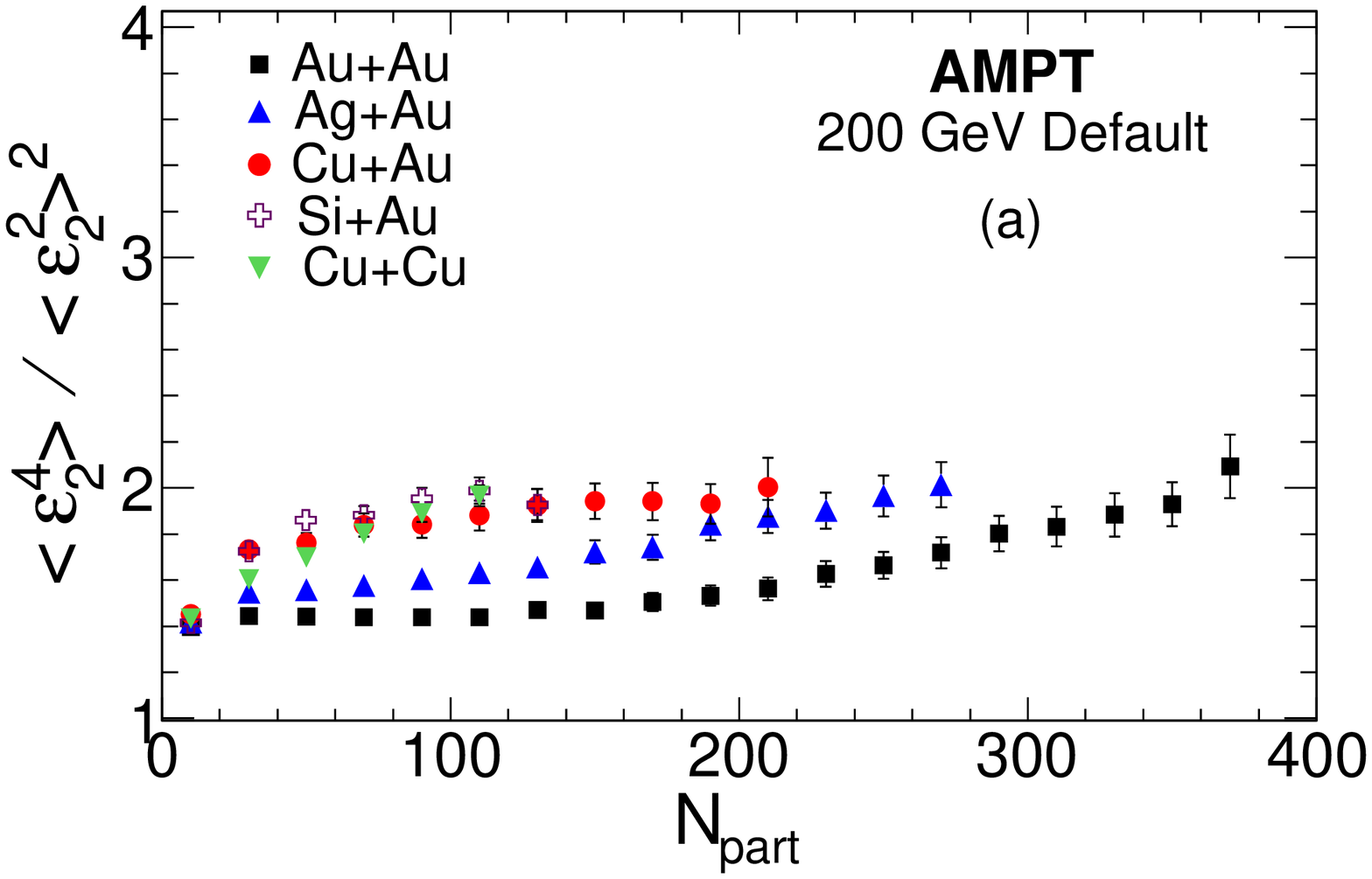}
\includegraphics[scale=0.32]{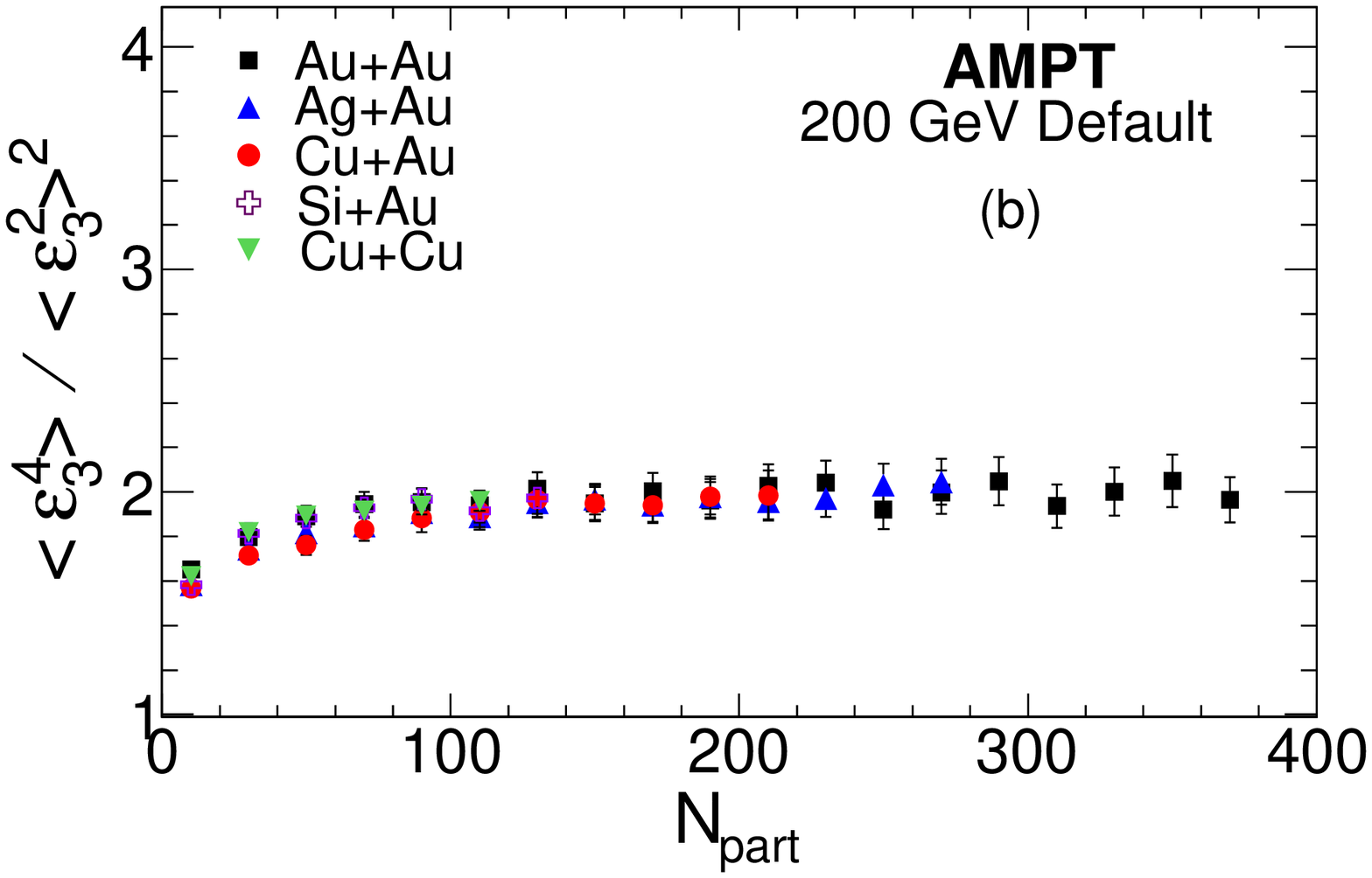}
\caption{(Color online) $\langle\varepsilon_n^4\rangle/\langle\varepsilon_n^2\rangle^2$, 
with $n=2,3$, versus $N_{\rm {part}}$ for various heavy-ion collisions at $\sqrt{s_{\rm {NN}}}$ = 200 GeV
using the AMPT model.}
\label{fig3}
\end{center}
\eef

\bef
\begin{center}
\includegraphics[scale=0.32]{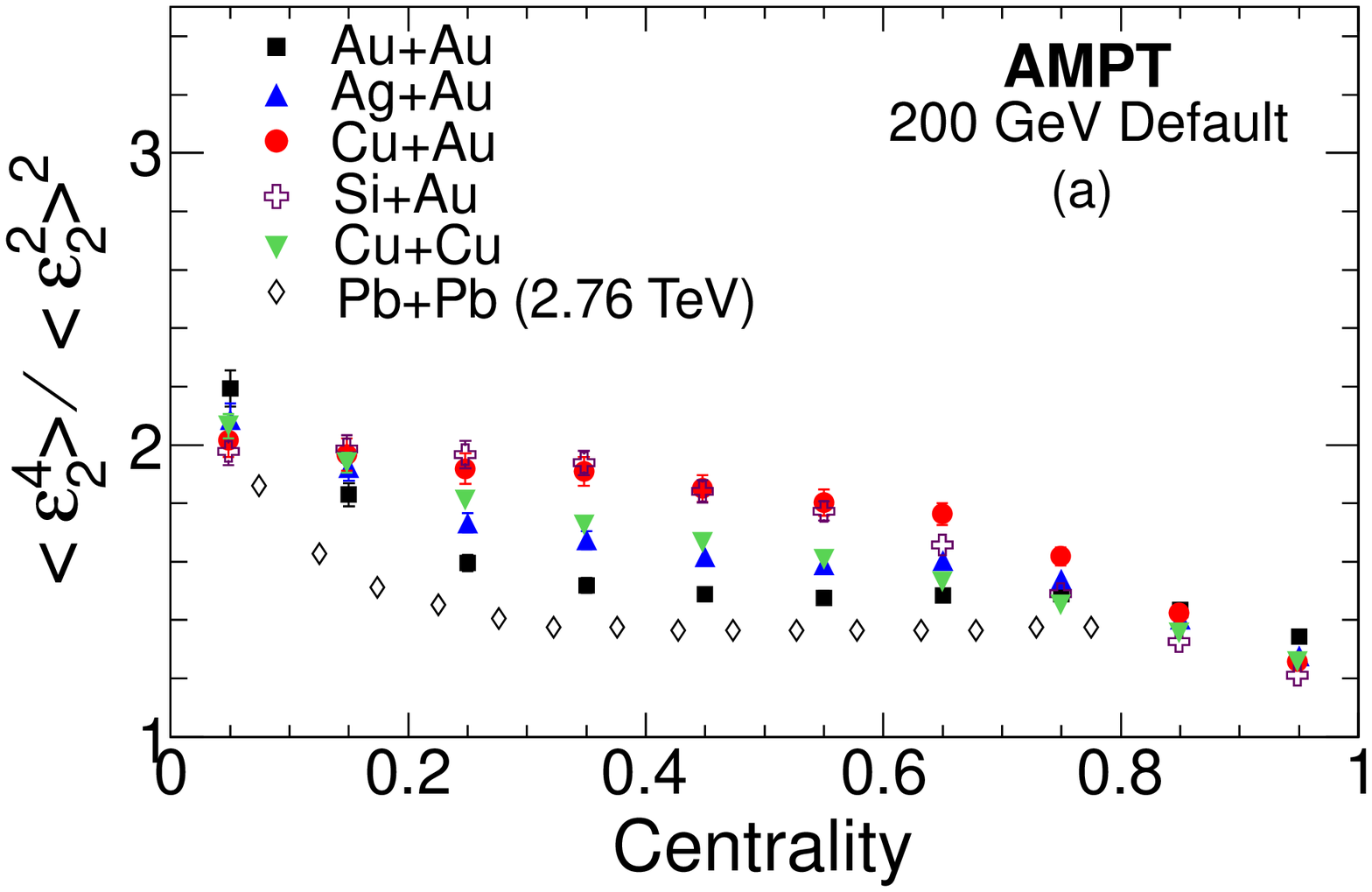}
\includegraphics[scale=0.32]{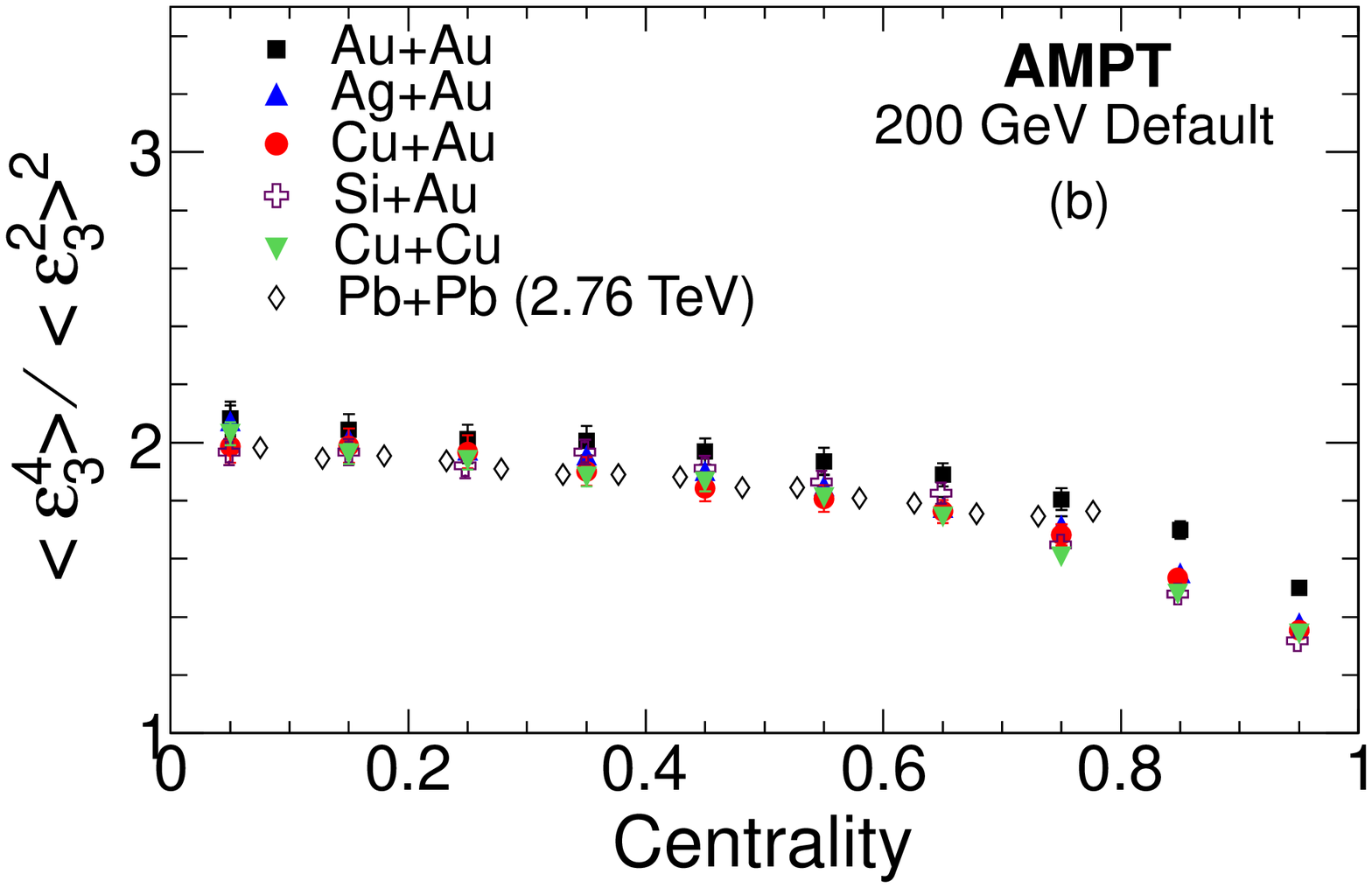}
\caption{ (Color online) $\langle\varepsilon_n^4\rangle/\langle\varepsilon_n^2\rangle^2$, 
with $n=2,3$, versus centrality for various heavy-ion collisions at $\sqrt{s_{\rm {NN}}}$ = 200 GeV. The 
Pb+Pb results corresponds to Glauber model simulations from Ref.~\cite{bhalerao} at $\sqrt{s_{\rm {NN}}}$ = 2.76 TeV.}
\label{fig4}
\end{center}
\eef

Figure~\ref{fig4} shows the same results as in Fig.~\ref{fig3} but as a function of fraction of collision
centrality. Where 0 corresponds to the most central collisions and 1 corresponds to the most peripheral collisions.
In addition we also show for comparison results from a Glauber model simulation for Pb+Pb collisions at
$\sqrt{s_{\rm {NN}}}$ = 2.76 TeV from Ref.~\cite{bhalerao}.
Since the colliding ion species have different maximum $N_{\rm {part}}$ values, it is desirable to study how 
the fluctuations in $\ecc$ and $\tria$ vary for the case of same fraction of collision centrality. 
From Fig.~\ref{fig4} one observes that although the fluctuations in $\tria$ is still similar for all the
colliding ion species studied, those for $\ecc$ now starts showing the ion size dependence more clearly.
For a given fraction of collision centrality the fluctuations in $\ecc$ starts to increase as the
colliding ion species size decreases, except for the most central and peripheral collisions.
These results show that a systematic study of heavy-ion collisions with 
colliding system sizes ranging between Cu+Cu and Au+Au may provide a better handle on understanding $\ecc$ fluctuations
and its relevance to $v_{\rm 2}$ fluctuations in the experimental measurements.

Collective flow is mostly driven by the initial spatial anisotropy, hence it is expected that $v_{\rm 2}$ and
$v_{\rm 3}$ should be proportional to $\ecc$ and $\tria$ respectively in high energy heavy-ion collisions.
Figure~\ref{fig5} shows $\langle v_{2} \rangle$ versus $\langle \ecc \rangle$ and $\langle v_{3} \rangle$ 
versus $\langle \tria \rangle$
from the AMPT model at midrapidity for 80 $<$ $N_{\rm {part}}$ $<$ 120 in heavy-ion collisions studied 
at $\sqrt{s_{\rm {NN}}}$ = 200 GeV.
\bef
\begin{center}
\includegraphics[scale=0.32]{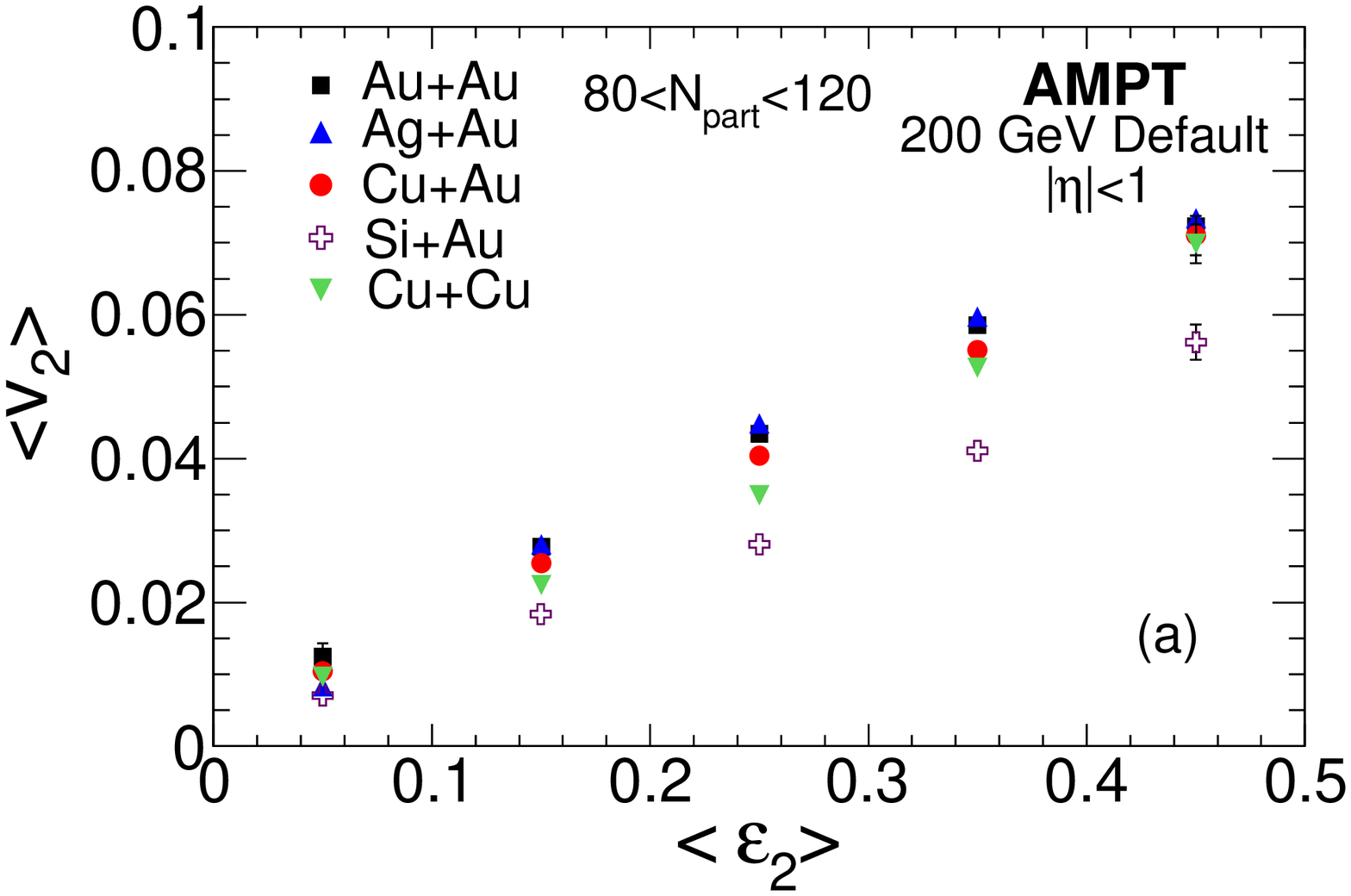}
\includegraphics[scale=0.32]{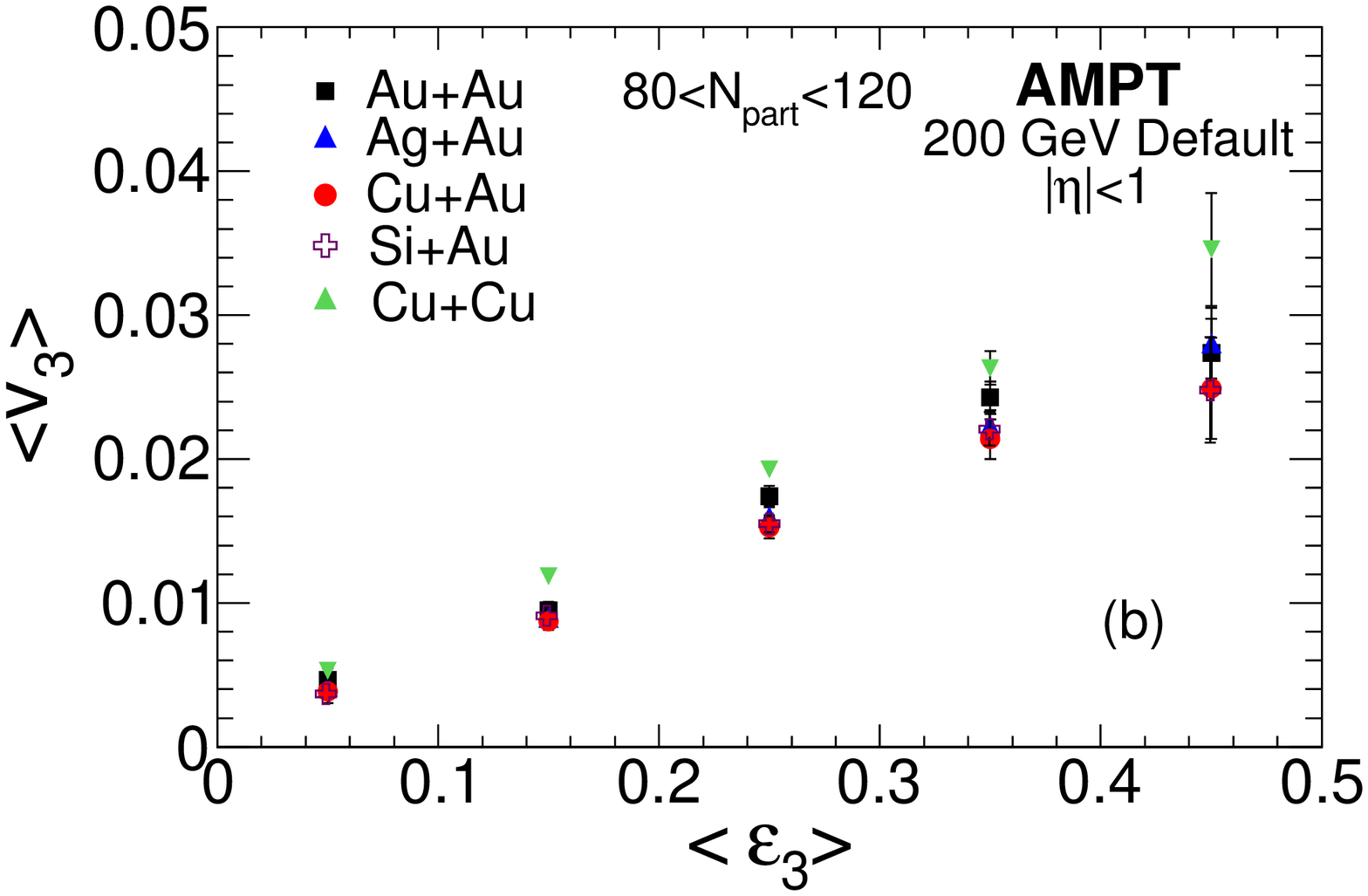}
\caption{(Color online) $\langle v_{\rm 2} \rangle $ (a) and $\langle v_{\rm 3} \rangle $  (b) of charged particles 
at midrapidity for 80 $<$ $N_{\rm {part}}$ $<$ 120 versus $\langle \ecc \rangle$ and $\langle \tria \rangle$ 
respectively. The results shown are for various species of heavy-ion collisions at $\sqrt{s_{\rm {NN}}}$ 
= 200 GeV from AMPT model. The errors shown are statistical.}
\label{fig5}
\end{center}
\eef
We observe that indeed $\langle v_{2} \rangle$ and $\langle v_{3} \rangle$ are proportional to $\langle \ecc \rangle$ 
and $\langle \tria \rangle$ respectively for 
all the collision systems studied. However for a given value of $\langle \ecc \rangle$, the observed  $\langle v_{\rm 2} \rangle$ 
decreases with a decrease in the colliding system size. This could be due to larger fluctuations
in $\ecc$ for the smaller systems compared to the larger colliding systems. However there is no such
observed differences for $\langle v_{\rm 3} \rangle$ versus $\langle \tria \rangle$. This is consistent with 
not much difference in fluctuations of $\tria$ for various colliding ions.

\bef
\begin{center}
\includegraphics[scale=0.32]{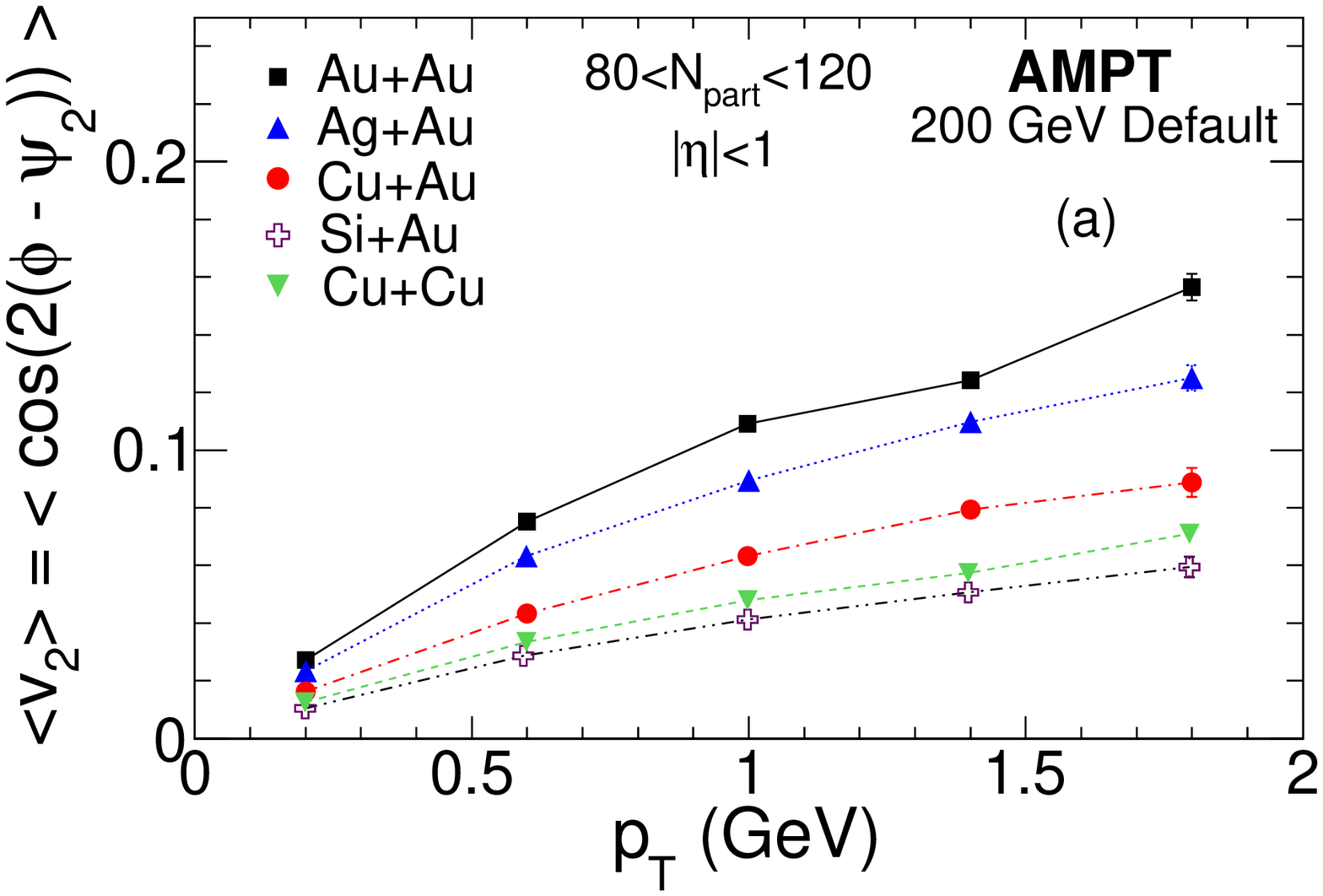}
\includegraphics[scale=0.32]{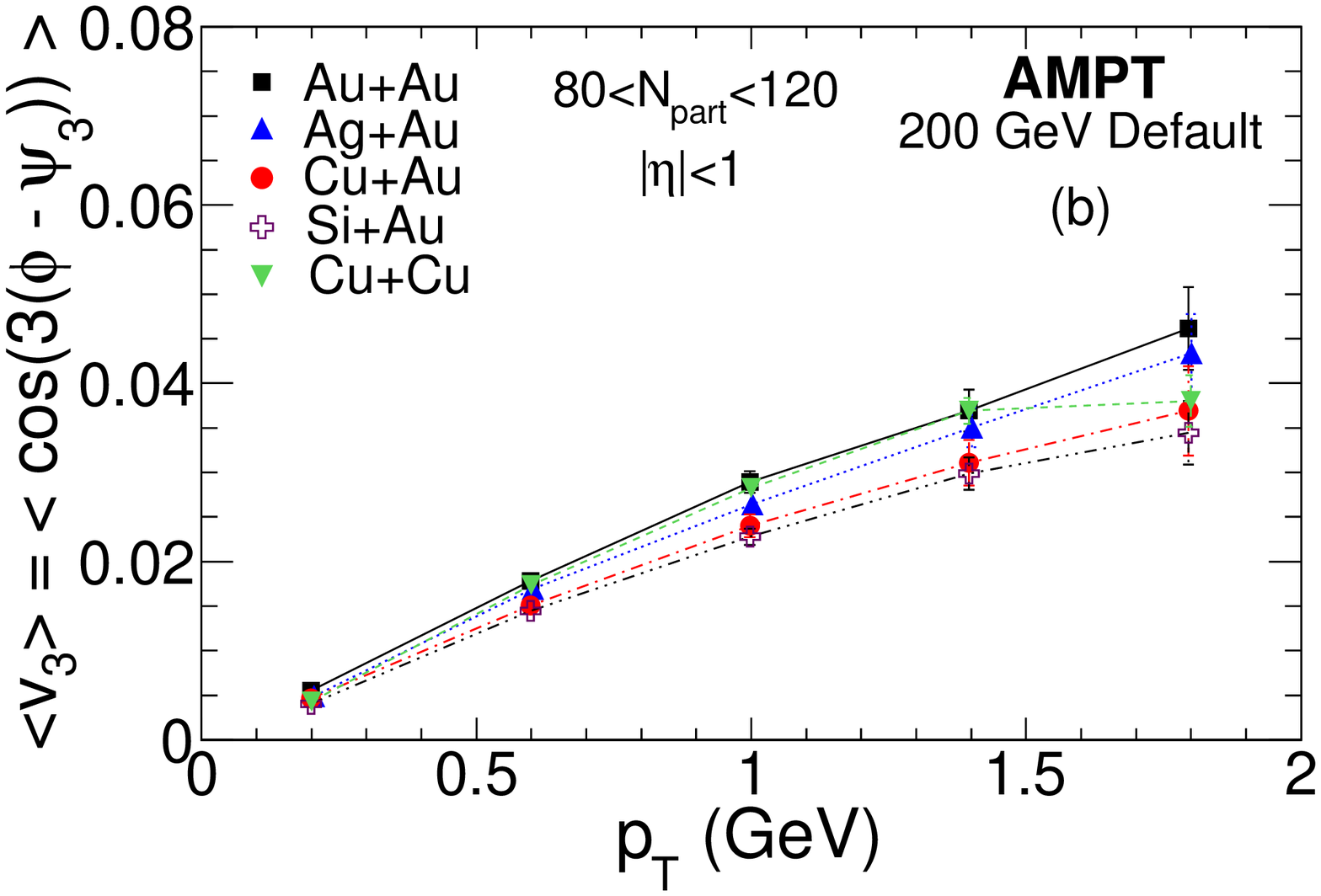}
\caption{(Color online) $v_{\rm 2}$ (a)  and $v_{\rm 3}$ (b) for charged particles as a function of transverse momentum ($p_{\mathrm T}$) 
at midrapidity for 80 $<$ $N_{\rm {part}}$ $<$ 120  for various colliding systems
at $\sqrt{s_{\mathrm {NN}}}$ = 200 from AMPT model. The errors are statistical.}
\label{fig6}
\end{center}
\eef
Finally in Fig.~\ref{fig6}(a) we show the $v_{\rm 2}$ for charged particles at midrapidity
for mid-central collisions as a function of transverse momentum ($p_{\rm T}$) for various
colliding systems at $\sqrt{s_{\rm NN}}$ = 200 GeV from the AMPT model. For the centrality 
range studied we see a clear dependence of $v_{\rm 2}$ on the colliding system size.
It increases with an increase in the colliding ion size. This is consistent with the results
on variation of $\langle v_{\rm 2} \rangle$ with $\langle \ecc \rangle$ shown in Fig.~\ref{fig5} 
and the fluctuations in $\ecc$ shown in Figs.~\ref{fig3} and ~\ref{fig4}. In Fig.~\ref{fig6}(b) the $v_{\rm 3}$ versus $p_{\rm T}$ 
shows a much smaller dependence on colliding system size compared to the above case.

\section{SUMMARY}
In view of the proposed asymmetric heavy-ion collision program at RHIC in 2012, 
we have presented an AMPT model based study of $\ecc$,
$v_{2}$, $\tria$, $v_{3}$ and fluctuations in $\ecc$ and $\tria$ for various asymmetric 
colliding ion species of Ag+Au, Cu+Au and Si+Au at $\sqrt{s_{NN}}$ = 200 GeV.
These results are presented as a function of the number of participating nucleons,
fraction of collision centrality and transverse momentum. They are compared to
results from symmetric colliding systems of Au+Au and Cu+Cu collisions.

Measurable signals of each of the above observables are found for all the
colliding systems studied. We find that while $\ecc$ and its fluctuations, 
for a given number of participating nucleons or fraction of collision centrality, 
are highly dependent on  the colliding ion type those for $\tria$ and its
fluctuations are very similar for all the colliding species studied. 
These results are reflected in the experimental observables such as $v_{2}$ and $v_{3}$.
For the same $\ecc$ at midrapidity and for mid-central collisions, the 
proportionality constant between $v_{2}$ and $\ecc$ seems to depend on the
colliding system size. On the other hand, $v_{2}$ vs. $p_{T}$ decreases as the colliding 
system size decreases for the collision centrality range studied. However 
no such large sensitivity to colliding ion type on $v_{\rm 3}$, $\tria$ and fluctuations 
in $\tria$ are observed from our AMPT model based study.
Our study thus indicates that asymmetric heavy-ion collisions can be used to 
constrain models dealing with flow fluctuations in heavy-ion collisions but with
greater sensitivity for $v_{2}$ related observables than for  $v_{\rm 3}$.

\noindent{\bf Acknowledgments}\\
This work is supported by
the DAE-BRNS project sanction No. 2010/21/15-BRNS/2026.

\end{document}